\documentclass[12pt, centerh1]{article} 
\usepackage{amssymb,color,multirow}

\usepackage{amsfonts}
\usepackage{natbib}
\usepackage{colonequals} 
\usepackage{amsmath}
\usepackage{epsfig}

\usepackage{enumerate}

\newcommand{\vecxc}{\mathbf{X}}

\newcommand{\vecx}{\mathbf{x}}

\newcommand{\load}{\mathbf{\Lambda}}
\newcommand{\noisev}{\mathbf\Psi}

\newcommand{\ident}{\mathbf{I}}

\newcommand{\vecmu}{\mbox{\boldmath$\mu$}}

\newcommand{\matb}{\mbox{\boldmath$\beta$}}

\newcommand{\matu}{\mathbf{u}}

\newcommand{\matsig}{\mathbf{\Sigma}}

\newcommand{\varthet}{\boldsymbol{\vartheta}}
\newcommand{\vecthet}{\mbox{\boldmath$\vartheta$}}
\newenvironment{theorem}[1][Theorem]{\begin{trivlist}
\item[\hskip \labelsep {\bfseries #1}]}{\end{trivlist}}
\newenvironment{pta}[1][Part a]{\begin{trivlist}
\item[\hskip \labelsep {\bfseries #1}]}{\end{trivlist}}
\newenvironment{ptb}[1][Part b]{\begin{trivlist}
\item[\hskip \labelsep {\bfseries #1}]}{\end{trivlist}}
\newenvironment{ptc}[1][Part c]{\begin{trivlist}
\item[\hskip \labelsep {\bfseries #1}]}{\end{trivlist}}
\newenvironment{ptd}[1][Part d]{\begin{trivlist}
\item[\hskip \labelsep {\bfseries #1}]}{\end{trivlist}}
\title{An Adaptive LASSO-Penalized BIC}

\author{Sakyajit Bhattacharya and Paul D.~McNicholas}
\author{Sakyajit Bhattacharya and Paul D.\ McNicholas}
\date{Dept.\ of Mathematics and Statistics, University of Guelph, Canada.}

\begin{document}

\maketitle

\begin{abstract}
Mixture models are becoming a popular tool for the clustering and classification of high-dimensional data. In such high dimensional applications, model selection is problematic. The Bayesian information criterion, which is popular in lower dimensional applications, tends to underestimate the true number of components in high dimensions. We introduce an adaptive LASSO-penalized BIC (ALPBIC) to mitigate this problem. This efficacy of the ALPBIC is illustrated via applications of parsimonious mixtures of factor analyzers. The selection of the best model by ALPBIC is shown to be consistent with increasing numbers of observations based on simulated and real data analyses.
\end{abstract}

\section{Introduction}
\label{se:intro}
The idiom `model-based clustering' refers to the application of mixture models for clustering. The first such application of mixture models for clustering dates back at least fifty years \citep{wolfe63} and until relatively recently, the Gaussian mixture model has dominated.
Consider $n$ realizations $\vecx_1, \vecx_2,..., \vecx_n$ of a $p$-dimensional random variable $\vecxc$ that follows a $G$-component finite Gaussian mixture model. The likelihood is given by
$\mathcal{L}(\varthet)
= \prod_{i=1}^n\sum_{g=1}^G \pi_g\phi(\vecx_i\mid\vecmu_g,\matsig_g)$,
where $\pi_g>0$, with $\sum_{g=1}^G\pi_g=1$, are the mixing proportions, $\phi(\vecx\mid\vecmu_g,\matsig_g)$ is the multivariate Gaussian density with mean $\vecmu_g$ and covariance matrix $\matsig_g$, and $\varthet$ denotes the model parameters. 
In model-based clustering applications, the expectation-maximization (EM) algorithm \citep{dempster77} is most often used for parameter estimation. The EM algorithm is based the complete-data likelihood, i.e., the likelihood of the observed data plus the unobserved data. In the context of model-based clustering using a Gaussian mixture model, the component labels are the unobserved data and the complete-data log-likelihood is given by: $\mathcal{L}_{\text{c}}(\varthet)=\prod_{i=1}^n\prod_{g=1}^G[\pi_g\phi(\vecx_i\mid\vecmu_g,\matsig_g)]^{z_{ig}}$, where the $z_{ig}$ denote the component membership labels so that $z_{ig}=1$ if observation~$i$ is in component~$g$ and $z_{ig}=0$ otherwise. 

Typically, the number of components $G$ and is selected using a model selection criterion. The Bayesian information criterion \citep[BIC;][]{schwarz78} is by far the most popular choice \citep[cf.][]{kass1,kass2,dasgupta98} and is given by: $\text{BIC}=2\log \mathcal{L}(\hat\varthet)-\rho\log n$, where $\hat\varthet$ is the MLE of $\varthet$ and $\rho$ is the number of free parameters. In addition to the selection of $G$, the BIC may be also be used to select the number of latent variables where relevant. 
In high-dimensional applications, the BIC tends to underestimate $G$. \cite{bhattacharya12} addressed the problem by introducing a LASSO-penalized BIC (LPBIC), where the parameter estimates and the model selection criterion were derived by maximizing a penalized log-likelihood with the penalty akin to the LASSO \citep{tibshirani}. The LPBIC is defined as 
\begin{equation}
\text{LPBIC}= 2\log\mathcal{L}(\hat\varthet \mid \vecx)-\tilde \rho \log n-\frac{2n\lambda_n}{G}\sum_{g=1}^G \sum_{j=1}^{p_{g}}\left[|\hat\mu_{gj}|+\frac{\left(I(\hat{\vecmu}_g)^{-1}\right)_{jj}}{|\hat\mu_{gj}|}-\mbox{sign}\left(\hat\mu_{gj}\right)\right], \label{eq:lpbic}
\end{equation} 
where $\tilde \rho$ is the number of estimated parameters which are non-zero, $p_g$ is the number of non-zero elements in $\vecmu_g$, $\hat\vecmu_g$ is the MLE derived by maximizing  the penalized log-likelihood $\log\mathcal{L}_{\text{pen}}$, $I(\hat\vecmu_g)$ is the unit information matrix at $\hat\vecmu_g$, and $\lambda_n$ is the tuning parameter. \cite{bhattacharya12} give several examples where the LPBIC outperforms the BIC in high-dimensional settings; however, the LPBIC is not without its drawbacks. Because the criterion is derived by using a LASSO-penalized log-likelihood, it suffers from the limitations of the LASSO. More particularly, the criterion does not satisfy the oracle properties \citep[cf.][]{fan}, i.e., it cannot simultaneously satisfy consistency, sparsity, and asymptotic normality. 

As an alternative, we propose a model selection criterion that is derived via a penalized log-likelihood, where the penalty is akin to the adaptive LASSO \citep{zou}. The adaptive LASSO was proposed to mitigate the consistency problems of the LASSO and \cite{zou} considered the case where $p$ is fixed.   Further progress on establishing the consistency of adaptive LASSO in a sparse high-dimensional linear regression set-up has been achieved by \cite{huang}. Recently, \cite{zhou} proved consistency in high-dimensional settings for Gaussian graphical modelling. Because these works mainly concern a sparse high-dimensional linear regression set-up, we can easily draw parallels in mixture modelling. For example, \cite{pan2} used the adaptive LASSO procedure in mixture models but their model selection criterion did not put any constraint on the tuning parameter and the weights used in the penalty term, thus making the consistency of the criterion questionable. \cite{pan} and \cite{pan2} used two different penalization approaches, one with the LASSO and another with the adaptive LASSO; however, they used the same model selection criteria for both. The work presented herein shows that the model selection criteria derived from the adaptive LASSO approach should differ significantly from that derived from the conventional LASSO approach. 

While deriving the MLE of the unknown parameters, we maximize the penalized log likelihood
\begin{equation}\label{eq:penlike}
\mathcal{L}_{\text{pen}}(\varthet \mid \vecx)=\log\mathcal{L}(\varthet \mid \vecx)- \sum_{g=1}^G \pi_g \sum_{j=1}^p \varphi(\mu_{gj}),
\end{equation}
where $\varphi(\mu_{gj})$ is an adaptive LASSO-like penalty. In particular, $\varphi(\mu_{gj})= n\lambda_n  w_{gj} \mid \mu_{gj} \mid $, where $\mu_{gj}$ is the $j$th element in $\vecmu_g$, the $w_{gj}$ are the weights, and $\lambda_n$ is a tuning parameter that depends on~$n$. 
The existing literature on adaptive LASSO suggests different choices for the weights $w_{gj}$. The present paper uses the choice suggested by \cite{zou}, i.e., that $w_{gj}= \mid \tilde\mu_{gj} \mid^{-\gamma}$ for some $\gamma >0$, where $\tilde\mu_{gj}$ is a $\sqrt{n}$-consistent estimate of $\mu_{gj}$ found by maximizing the (unpenalized) likelihood. 
In our theoretical discussions, we take $\gamma\in[0,1]$. When illustrating our method (Section~\ref{se:data}), we set $\gamma=1$, which is the most popular choice. \cite{zou} and \cite{pan2} use weights $w_{gj}= \mid \tilde\mu_{gj} \mid^{-1}$, but other choices may work better. For example, \cite{zou} noted that in high-dimensional cases an $L_2$-penalized estimator can be a better choice. \cite{zhou} proposed $w_{gj}= \mbox{max} \left(\mid \tilde\mu_{gj} \mid^{-1}, 1\right)$. Also, \cite{meinshausen} and \cite{zhang} pointed out that estimation by maximizing the unpenalized likelihood is not always feasible in a high dimensions. This is particularly true for a regression problem, especially when there exists strong collinearity among the covariates. 
 
Herein, we develop and illustrate our ALPBIC for parsimonious mixtures of factor analyzers models. We choose these models because they are well suited to the analysis of high dimensional data and because model selection is more involved than just selecting the number of components. 
\cite{ghahramani97} introduced a mixture of factor analyzers model, which was further developed by \cite{tipping99b} and \cite{mclachlan00a}. Through imposing constraints on the covariance structure, \cite{mcnicholas08,mcnicholas10d} develop a family of parsimonious Gaussian mixture models (PGMMs) based on the mixture of factor analyzers model. The mixture of factor analyzers model has the density of a Gaussian mixture model with $\matsig_g=\load_g\load_g'+\noisev_g$, where $\load_g$ is a $p\times q$ matrix of factor loadings, $\noisev_g$ is a $p\times p$ diagonal matrix, $p$ is the dimensionality of the data, and $q$ is the dimensionality of the latent factors. The PGMM family arises by imposing, or not, the constraints $\load_g=\load$, $\noisev_g=\noisev$, and $\noisev_g=\psi_g\ident$ on the component covariance matrix $\matsig_g$. 
We show that when $p$ is large and $n \longrightarrow \infty$, ALPBIC generally outperforms the LPBIC and the BIC in the PGMM setting in terms of consistently selecting the right number of components as well as more accurately classifying observations into components.
 


\section{Methodology}\label{se:method}
Recall that we observe $\vecx=\left(\vecx_1, \vecx_2,..., \vecx_n \right)'$ with $f(\vecx\mid\varthet)= \sum_{g=1}^G \pi_g\phi(\vecx\mid\vecmu_g,\matsig_g)$. Now, denote 
$\varphi(\vecmu)= \sum_{g=1}^G \pi_g \sum_{j=1}^p w_{gj}|\mu_{gj}|$ and so
$$\mathcal{L}_{\text{pen}}(\varthet \mid \vecx)
=\log\mathcal{L}(\varthet \mid \vecx)-n\lambda_n\sum_{g=1}^G \pi_g \sum_{j=1}^p \varphi_{gj}(\vecmu)
=\log\mathcal{L}(\varthet \mid \vecx)-n\lambda_n\varphi(\vecmu).$$
Here we make two assumptions. Firstly, because the penalty function is non-concave and singular at the origin, it can be locally approximated by a quadratic function as suggested by \cite{fan}. The parameters are estimated by successive iterations. Suppose $\vecmu^{(m)}$ is the estimate of of $\vecmu$ after $m$ iterations. Then, the penalty can be locally approximated as 
\begin{equation}
 \varphi(\vecmu)  \approx \,  
 n\lambda_n \sum_{g=1}^G \pi_g \sum_{j=1}^{p_g} w_{gj}|\mu^{(m)}_{gj}|+\frac{1}{2}w_{gj}\frac{\mbox{sign}(\mu^{(m)}_{gj})}{\mu^{(m)}_{gj}}(\mu^2_{gj}-{\mu_{gj}^{(m)}}^2),\label{eq:penalty}
\end{equation}  
where $p_g$ is the number of non-zero elements in $\vecmu_g$.
Secondly, we assume that the marginal distribution of the mixing proportions $\left(\pi_1,\pi_2,...,\pi_g\right)$ is uniform on the simplex and $\vecmu_g \sim \mathcal{N}\left(\hat\vecmu_g, I(\hat\vecmu_g)^{-1}\right)$, for $g=1,2,...,G$, where $\hat\vecmu_g$ is the MLE derived by maximizing  the penalized likelihood $\mathcal{L}_{\text{pen}}$ and $I(\hat\vecmu_g)$ is the unit information matrix at $\hat\vecmu_g$. 

With these assumptions, parameter estimation is carried out using an alternating expectation-conditional maximization (AECM) algorithm \citep{meng97}. The AECM algorithm allows different specification of the complete-data at each stage. At the first stage, $\mathbf{\pi}$ and $\vecmu$ are estimated and at the second stage, the elements of $\matsig$ are estimated. The two stages are alternated iteratively until convergence; see \cite{mcnicholas08} for extensive details on the application of the AECM algorithm for the PGMM models. 
 
At the first stage of the AECM algorithm, the complete-data comprise the observed $\vecx_i$ and the component labels $\mathbf{z}_i$. The estimation of $\pi_g$ involves complexity for a penalized likelihood. However, \cite{bhattacharya12} note that for practical applications, the difference between the analytical estimate and the estimate derived by the conventional EM algorithm for an unpenalized likelihood is negligible. Hence the mixing proportions are estimated via $\hat\pi_g= {\sum_{i=1}^n \hat z_{ig}}/{n}$, where $$\hat z_{ig}=\frac{\hat\pi_g \phi(\vecx_i \mid \hat\vecmu_g,\hat\matsig_g)}{\sum_{j=1}^G \hat\pi_j\phi(\vecx_i \mid \hat\vecmu_g,\hat\matsig_g)}$$ denotes the expected value of $Z_{ig}$ conditional on the current parameter estimates.
For the mean parameters, we differentiate the expected value of the complete-data log-likelihood with respect to $\vecmu_g$ and equate the result to zero, i.e., 
\begin{equation}\label{muscore}
\hat\matsig_g^{-1} \sum_{i=1}^n \hat z_{ig}(\vecx_i-\vecmu_g)-n\lambda_n \hat\pi_g\matb_g=0,\end{equation} where $\matb_g$ is a vector with $p$ elements, its $j$th element being $w_{gj}\mbox{sign}(\mu_{gj})$.
Now, replacing $\vecmu_g$ by~$\hat{\vecmu}_g$, \eqref{muscore} can be written 
\begin{equation}
 \tilde\mu_{gj}-\hat\mu_{gj}-\lambda_n \left( \hat\matsig_g \hat{\mathbf{\Lambda}}_g\right)_j = 0,\label{eq:em1} 
\end{equation} 
  where $\tilde\mu_{gj}=\sum_{i=1}^n\hat{z}_{ig}x_{ig}/\sum_{i=1}^n\hat{z}_{ig}$ is the update of $\mu_{gj}$ if no penalty term were involved. 
Furthermore, imposing an adaptive LASSO penalty on the mean components implies that the estimate of $\mu_{gj}$ for a non-penalized case, i.e. $\tilde\mu_{gj}$, is shrunken towards zero, leading to a sparse solution. Therefore, the new estimate is either zero or takes a value lying between zero and $\tilde\mu_{gj}$. In other words, either $\hat\mu_{gj}=0$ or $\mbox{sign}\left(\hat\mu_{gj}\right)=\mbox{sign}\left(\tilde\mu_{gj}\right)= \mbox{sign}\left(w_{gj}\tilde\mu_{gj}\right)$ because $w_{gj}$ is positive. Thus, if $\hat\mu_{gj} \neq 0$, then 
\begin{equation}
\mid \tilde\mu_{gj}-\hat\mu_{gj} \mid < \mid \tilde\mu_{gj} \mid . \label{eq:em2}
\end{equation}
Using (\ref{eq:em1}) in (\ref{eq:em2}) and the sparsity described above, we get 
$\lambda_n < |\tilde\mu_{gj}|/\big|\left( \hat\matsig_g \tilde{\matb}_g\right)_j\big|$ and $\mbox{sign}(\tilde\mu_{gj})=\mbox{sign}\left(\hat\matsig_g \tilde{\matb}_g\right)_j$ because $\lambda_n$ is non-negative, where $\tilde{\matb}_{gj}= w_{gj}\mbox{sign}(\tilde\mu_{gj})$. Equation~(\ref{eq:em1}) and the above arguments lead to the following estimate of $\mu_{gj}$: 
\begin{eqnarray*}
\hat{\mu}_{gj}= \mbox{sign}(\tilde\mu_{gj})\left[|\tilde\mu_{gj}|-\lambda_n \big|\left(\hat\matsig_g\tilde{\mathbf{\Lambda}}_g\right)_j\big|\right]_{+},
\end{eqnarray*} 
where, for any $\alpha$, $\alpha_{+}=\alpha$ if $\alpha >0$ and $\alpha_{+}=0$ otherwise.  

At the second stage of the AECM algorithm, the complete-data comprise the observed $\vecx_i$, the group labels $\mathbf{z}_i$, and the unobserved latent factors $\mathbf{u}_{ig}$. At this stage, we estimate the elements of the covariance matrix under the PGMM set-up and details are given by \cite{mcnicholas08,mcnicholas10d}. 

To derive a model selection criterion from the penalized log-likelihood we apply the same method as used by \cite{bhattacharya12}. We maximize (\ref{eq:penlike}). Using~(\ref{eq:penalty}), the second term of (\ref{eq:penlike}) becomes
$$\frac{\lambda_n}{G}\sum_{g=1}^G \sum_{j=1}^{p_{g}}w_{gj}\left[|\hat\mu_{gj}|+\frac{1}{2}\frac{\mbox{sign}(\hat\mu_{gj})}{\hat\mu_{gj}}(\mu^2_{gj}-\hat\mu^2_{gj})\right],$$ where $p_g$ is the number of non-zero mean components in component $g$.
We assume that for a given model the parameters for any two components are independent. Hence, using the Weak Law of Large Numbers, the penalized criterion is 
\begin{equation}
\text{ALPBIC}= 2\log\mathcal{L}(\hat\varthet \mid \vecx)-\tilde \rho \log n-\frac{2n\lambda_n}{G}\sum_{g=1}^G \sum_{j=1}^{p_{g}}w_{gj}\left[|\hat\mu_{gj}|+\frac{\left(I(\hat{\vecmu}_g)^{-1}\right)_{jj}}{|\hat\mu_{gj}|}-\mbox{sign}\left(\hat\mu_{gj}\right)\right], \label{eq:pbic}
\end{equation} 
where $\tilde \rho$ is the number of estimated parameters which are non-zero.
The derivation of \eqref{eq:pbic} is discussed in detail in Appendix~\ref{se:app1}.

\section{Asymptotic Properties}\label{asymp}

The ALPBIC is a consistent model selection criterion with conditions imposed on the tuning parameter $\lambda_n$. 
 To prove consistency we use some of the arguments proposed by \cite{khalili}, who originally studied the asymptotic behaviour within a low-dimensional framework; accordingly, we have modified their results for high-dimensional cases. Suppose the true parameter set $\varthet_0$ is decomposed as $(\varthet_{01}, \varthet_{02})$ such that $\varthet_{02}$ contains only the zero effects, and suppose any estimated parameter $\hat\varthet$ that is sufficiently close to $\varthet_0$ is likewise decomposed as $(\hat\varthet_1, \hat\varthet_2)$. Let $\vecmu$ be likewise decomposed as $(\vecmu_1, \vecmu_2)$. Then to satisfy consistency, the necessary criteria are $\text{P}(\hat\varthet_2 = \mathbf{0})\longrightarrow 1$ as $n \longrightarrow \infty$ and $\hat\varthet_1 \longrightarrow \varthet_{01} $ in probability. Thus, the criterion should choose as it would if the true number of clusters and the true parameters were known. Based on this idea, we assume some regularity conditions as stated below:
\begin{enumerate}[(i)]
\item $I(\varthet)$ is finite and positive-definite for all $\varthet$.  
\item For all $\varthet=\left(\theta_1,\theta_2,...,\theta_{\nu}\right)$ satisfying $|| \varthet-\varthet_0 || = \mathcal{O}\left(n^{-{1/2}}(\log n)^{1/2}\right) $, there exist finite real numbers $M_1$ and $M_2$ (possibly depending on $\varthet_0$) such that 
$$\sup_j\left| \frac{\partial \mathcal{L}(\varthet \mid \vecx)}{\partial \theta_j}\right| \leq M_1(\vecx) \quad \mbox{and} \quad \sup_{j,k}\left| \frac{\partial^2 \mathcal{L}(\varthet \mid \vecx)}{\partial \theta_j \partial\theta_k} \right| \leq M_2(\vecx),$$ with $$\int M_1(\vecx)d\vecx < \infty \qquad \text{and} \qquad \int M_2(\vecx)d\vecx < \infty.$$
\item For all $\varthet$ satisfying $|| \varthet-\varthet_0 || = \mathcal{O}\left(n^{-{1/2}}(\log n)^{1/2}\right)$, \begin{equation}\underset{g,j}\max\big\{I\left(\mu_{gj}\right)^{-1}_{jj}\big\} = \mathcal{O}\big(n^{-{1/2}}\big).\end{equation}\label{assump:3}
\end{enumerate}

The asymptotic properties of the ALPBIC are given in the following theorem.
\begin{theorem}
Suppose $G$ is known and let $p= \mathcal{O}(n^\kappa)$ for $\kappa \geq 0$. If $$n^{(2\kappa+1)/2}\lambda_n/(\log n)^{1/2} \longrightarrow 0 \quad\quad \text{and} \quad\quad n^{(\gamma+2\kappa+1)/2}\lambda_n/(\log n)^{1/2}=\mathcal{O}(1),$$ then we have the following:

\begin{pta}
There exists a local maximizer $\hat\varthet$ of ALPBIC for which $$\mid\mid \hat\varthet -\varthet_0 \mid\mid = \mathcal{O}\left(n^{-{1/2}}(\log n)^{1/2}\right).$$
\end{pta} 

\begin{ptb}
For such $\hat\varthet$, 
with probability tending to 1, $$\text{ALPBIC}\left(\hat\varthet_1, 
\hat\varthet_2 \right)-\text{ALPBIC}\left(\hat\varthet_1, \mathbf{0}\right) < 0 $$ provided $ \underset{g,j : \mu_{gj} \in \varthet_2}{\text{max}}\hat\mu_{g,j} \leq  \left(\log n/{n}\right)^{1/2}$ for sufficiently large $n$.
\end{ptb}

\begin{ptc}
$$\sqrt{n}\bigg[\mathbf{I}(\varthet_{01})\left(\hat\varthet_1-\varthet_{01}\right)+\lambda_n\sum_{g=1}^G\pi_g\sum_{j=1}^{p_{0g}} w_{gj}\text{sign}\left(\mu_{0gj}\right) \bigg]\overset{d}{\longrightarrow} \mathcal{N}\left(\mathbf{0}, \mathbf{I}(\varthet_{01})\right)$$ and $$\text{P}\left(\hat\vecmu_2=\mathbf{0}\right) \longrightarrow 1,$$ where $\mathbf{I}(\varthet_{01})$ is the Fisher information computed under the
reduced model when all zero effects are removed, and $p_{0g}$ is the number of non-zero elements in the $g$th component for the true parameter~$\varthet_0$.
\end{ptc}

\begin{ptd}
If $G$ is unknown and is estimated consistently by $\tilde G$ separately, then the results in Parts a, b, and c still hold when $\tilde G$ is subsequently used in the model selection procedure.
\end{ptd}
\end{theorem}

Part a of the above theorem guarantees the existence of a local maximizer of the parameter. Part b satisfies the sparsity condition. Part c satisfies the asymptotic normality condition. These results together constitute the oracle property. To prove that the oracle property is satisfied, we impose two conditions on the tuning parameters. As an aside, we can regard this as a sort of explanation as to why the LASSO cannot satisfy the oracle property, i.e., for the LASSO, $\gamma=0$ and so the two conditions cannot be simultaneously satisfied. Another observation worth noting is that, generally, for high-dimensional settings we take $\kappa > 0$ and we do so in the proof of the asymptotic properties in Appendix~\ref{app:b}. However, the same result happens with $\kappa=0$, i.e., for fixed~$p$, and the proof for fixed~$p$ in a regression setting is detailed by \cite{zou} and \cite{huang}. 


\section{Data Analysis}\label{se:data}
\subsection{Overview}
We analyze two data sets and compare the performance of ALPBIC with the LPBIC, the BIC, the AIC \citep{akaike}, and the CAIC \citep{bozdogan} for the PGMM family. The first one is a high-dimensional simulated data set and the second one is a real high-dimensional data set. For the first data set, we keep $p$ large and increase $n$ to study its effect on the consistency of the model selection criteria under consideration. The performance of a criterion is measured in terms of choosing the correct number of components and latent factors, and the correct covariance structure, as well as the resulting classification agreement. 
We use the adjusted Rand index \citep[ARI;][]{rand,hubert} to measure the latter. An ARI value of $1$ indicates perfect agreement and a value of $0$ would be expected under random classification. \cite{steinley04} provides extensive details on the ARI and its properties. Note that we set $\gamma=1$ for the ALPBIC and $\gamma=0$ for the LPBIC.

\subsection{Simulated Data}
We generate three $p$-dimensional Gaussian mixtures with $G=3$ components and $p=200$ variables. We use the following parameters in the component densities: $\vecmu_1=-5.5\mathbf{1}$, $\matsig_1$ isotropic; $\vecmu_2=2\mathbf{1}$, $\matsig_2$ diagonal; and $\vecmu_3=3\mathbf{1}$, $\matsig_3$ full with $(i,j)$th element $0.9^{\mid i-j \mid}$. Simulations are run in two steps. In the first step, we simulate three data sets, using $n\in\{40,100,200,500\}$ and setting the ratio of number of elements in the three groups to be $4:3:3$ in each case. The PGMM family was applied to these data for $G=1,\ldots,5$ and $q=1,\ldots,6$ using the ALPBIC, the LPBIC, and the BIC, respectively, for model selection. In the second step, we run 25 simulations for $p=200$, vary $n$ as before but also vary the relative size of clusters for each $n$. We then compare the performance of the ALPBIC with the LPBIC, the BIC, the AIC, and the CAIC in this scenario to study how the model selection criteria behave for varying $n$ and varying relative component sizes.

The results for the first step of experiment (Table~\ref{tab:sim1p}) show that the ALPBIC consistently chooses $G=3$, $q=4$, and the CUU covariance structure as $n$ gets larger. The LPBIC, on the other hand, consistently chooses $G=3$ but fails to consistently choose $q$ or the covariance structure.  The BIC, however, performs the worst: it fails to select the correct number of components. In fact, the BIC lives up to form by underestimating $G$ for each value of $n$; unsurprisingly, the value of $q$ is low. 
\begin{table}
\caption{\label{tab:sim1p} Best model chosen by ALPBIC, LPBIC and BIC for high-dimensional simulated data.}
\centering
\begin{tabular*}{1.00\textwidth}{@{\extracolsep{\fill}}llllllllllll}
\hline
   & \multicolumn{3}{c}{ALPBIC} && \multicolumn{3}{c}{LPBIC} && \multicolumn{3}{c}{BIC} \\
\cline{2-4} \cline{6-8} \cline{10-12}
 & $G$ & $q$ & Model  && $G$ & $q$ & Model && $G$ & $q$ & Model \\
\cline{2-4} \cline{6-8} \cline{10-12}
$n=40$   &$3$ &  $4$ & CUU    && $2$ & $1$ & CUC && $2$& $1$ & CCC \\ 
$n=100$ &$3$ &  $4$ & CUU  && $3$ & $3$ & CUC  && $2$ & $1$ & CUC  \\
$n=200$ &$3$ &  $4$ & CUU  && $3$ & $1$ & CUU  && $2$ & $1$ & CUC  \\
$n=500$ &$3$ &  $4$ & CUU && $3$ & $4$ & CUU && $2$ & $1$ & CUC  \\
\hline 
\end{tabular*}
\end{table}

\begin{figure}[!h]
\begin{center}
\begin{tabular}{cc}
\includegraphics[width=0.45\textwidth]{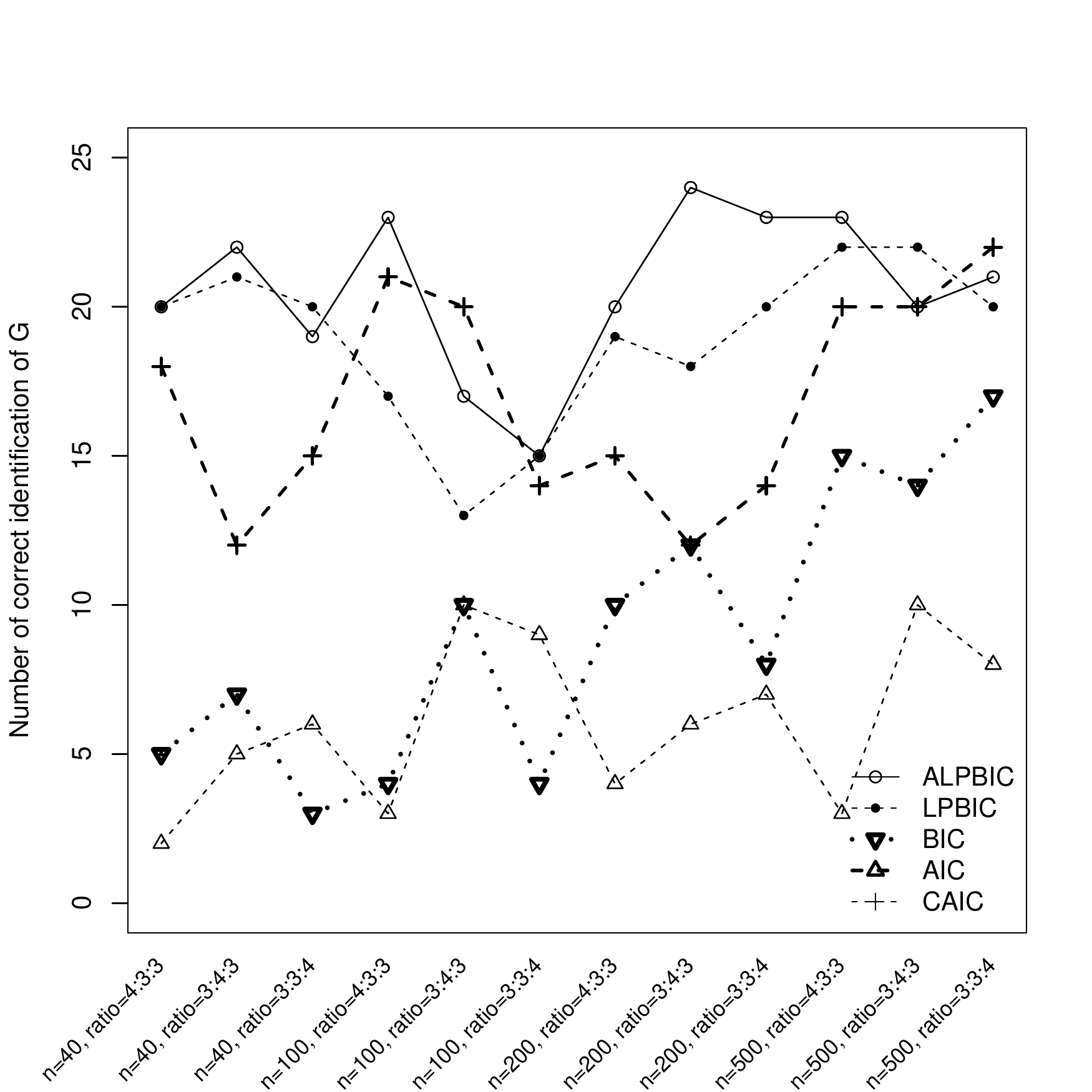} & \qquad \includegraphics[width=0.45\textwidth]{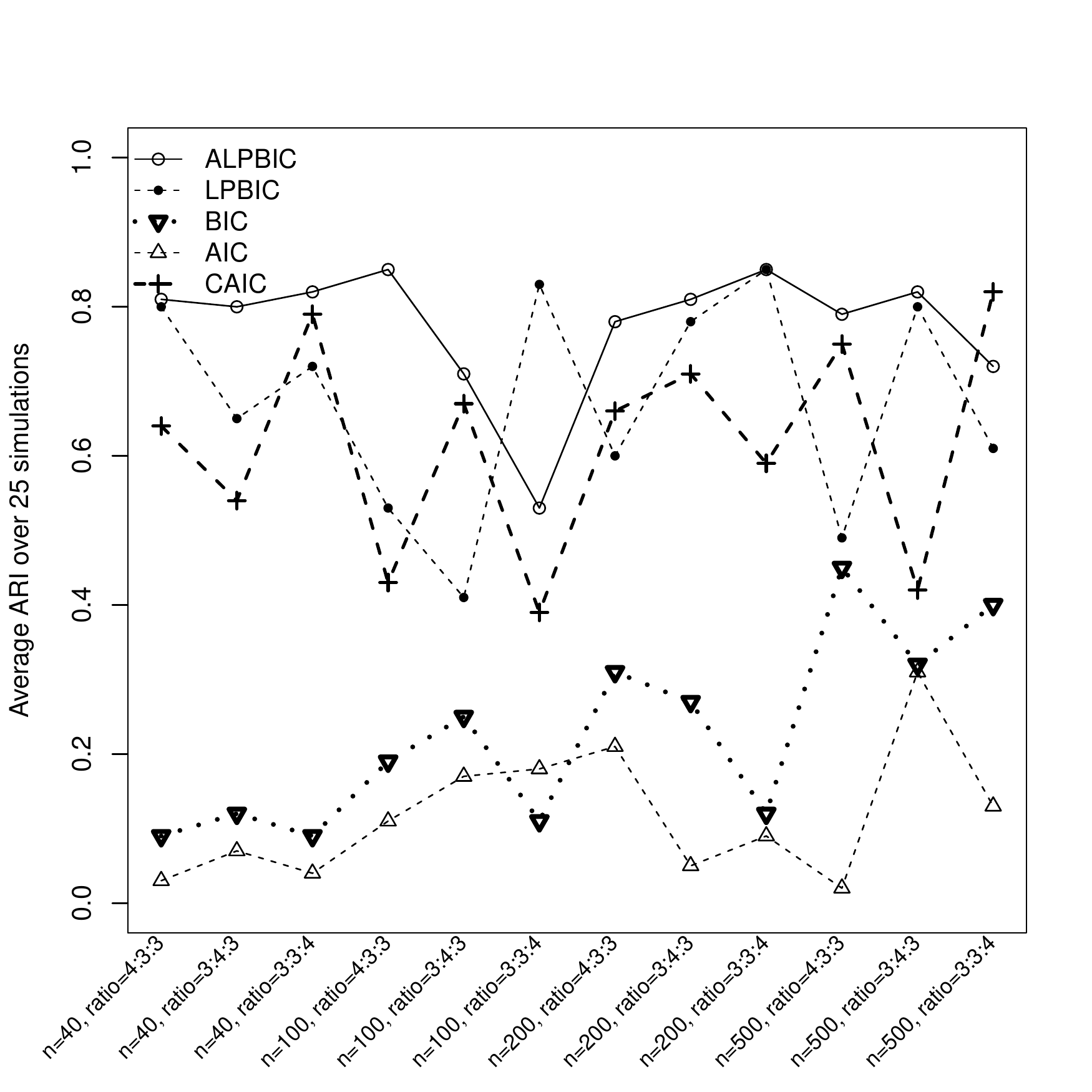}
\end{tabular}
\end{center}
\vspace{-0.2in}
\caption{Plot of the performance of ALPBIC, LPBIC, BIC, AIC AND CAIC for 25 simulations with varying $n$ and varying cluster size. The left-hand plot shows the selection of number of components (out of 25 simulations) by the five model selection criteria. The right-hand plot shows the average ARIs (over 25 simulations) of the models selected by the five model selection criteria.}\label{fi:sim1}
\end{figure}

In the second step, we vary $n$ as well as the relative sizes of the clusters, and generate 25 simulations for each set of parameter values. For each $n\in\{40,100,200,500\}$, three choices of relative component sizes are considered: 4:3:3, 3:4:3, and 3:3:4. In each of these 12 settings, the ratio of component sizes are chosen such that the proportion of data points is higher in one component and the remaining data are equally divided among the other two components. This is done to generate different covariance structures on varying sample sizes. We compare the performance of the ALPBIC, the LPBIC, the BIC, the AIC and the CAIC for each situation. First, we evaluate the ALPBIC on its ability to detect the correct number of components, comparing its performance with the other methods and using the same ranges of $G$ and $q$ as before. The left-hand plot of Figure~\ref{fi:sim1} shows that while $n$ and the relative cluster size vary, the ALPBIC is the most consistent in choosing the correct number of clusters (i.e., $G=3$). Out of 25 simulations in each case, the number of correct selections of $G$ is the highest for ALPBIC, closely followed by LPBIC and CAIC. ALPBIC has the highest number of correct selections of $G$ in 10 out of the 12 settings. The AIC and BIC are the poorest performers, but the latter's performance improves as $n$ increases. 

Because we do not know the correct $q$, it is difficult to study the performance of the ALPBIC in selecting $q$. However, the results show that ALPBIC chooses $q$ most consistently. Specifically, out of 25 simulations in each of the 12 settings, ALPBIC chooses $q=4$ an average of 72\% of time while LPBIC chooses $q=4$ an average of only 43\% of time. BIC and AIC show the lowest consistency in this respect; the average ratio is 12\% for both criteria. CAIC gives the most comparable performance to the ALPBIC, choosing $q=4$ an average of 61\% of time. Although we do not know the true value of $q$, $q=4$ seems a reasonable guess, based both on this experiment and the previous experiment (cf.\ Table~\ref{tab:sim1p}).

We also compare the classification performance of these five criteria for varying scenario. The right-hand plot of Figure~\ref{fi:sim1} shows the average classification accuracy (ARI) over 25 simulations for each setting for each criterion. The ALPBIC gives consistent classification results, with the average ARI usually lying between 0.7 and 0.85 (the exception is $n=100$, $3:3:4$). For one setting ($n=100$, $3:3:4$), the LPBIC outperforms the ALBIC, and for another, the CAIC does better ($n=500$, $3:3:4$). However, the ALBIC and CAIC generally return lower average ARI values and are less consistent (cf.\ Figure~\ref{fi:sim1}). The BIC and AIC, on the other hand, do not fluctuate much but have low average ARI values.


\subsection{Leukaemia data}

\cite{golub} presented data on two forms of acute leukaemia: acute lymphoblastic leukaemia (ALL) and acute myeloid leukaemia (AML). Affymetrix arrays were used to collect measurements for 7,129 genes on 72 tissues. There were a total of 47 ALL tissues and 25 with AML. \cite{mcnicholas10d} reduced the number of genes to 2,030 prior to analysis and we analyze these 2,030 genes using 20 different random starts for the initial $\hat z_{ig}$. We run our approach for $G\in\{1,2\}$ and $q=1,\ldots,6$.

\begin{table}
\vspace{.25cm}
\caption{\label{tab:luke1}Performance of ALPBIC, LPBIC, and BIC for the leukaemia data. The left-hand table gives the selected model and associated ARI for each criteria. The right-hand table gives cross-tabulations of predicted classifications versus true labels for the model selected by the ALPBIC and LPBIC, respectively.}
\centering
\begin{tabular*}{0.5\textwidth}{@{\extracolsep{\fill}}llllll}
\hline
  & Value & $G$ & $q$ & Model & ARI \\ 
\hline 
ALPBIC & $-388647$ & $2$ & $2$ & CUC &  0.51\\ 
LPBIC & $-391023$ & $2$ & $1$ & CUC & 0.47 \\   
BIC & $-400394$ & $1$ & $2$ & CCU &  0.29\\
\hline
\end{tabular*}
\hspace{1cm}
\begin{tabular*}{0.4\textwidth}{@{\extracolsep{\fill}}llllll}
\hline
 & \multicolumn{2}{c}{ALPBIC} && \multicolumn{2}{c}{LPBIC} \\
\cline{2-3} \cline{5-6}
& 1 & 2 && 1 & 2\\
\hline
\text{ALL} & 41 & 4 && 39 & 3 \\
\text{AML} & 6 &  21 && 8 & 22\\
\hline
\end{tabular*}
\end{table} 
Table~\ref{tab:luke1} summarizes the performance of the three model selection criteria. ALPBIC and LPBIC are comparable in selecting the model and classification agreement, as shown in Table~\ref{tab:luke1} (a). Both  ALPBIC and LPBIC  choose a CUC model with $G=2$ components. However, LPBIC selects $q=1$ and ALPBIC selects $q=2$ factors.
The BIC chooses a CCU model with $G=1$ component and $q=2$ factors. The ARI for the model chosen by the ALPBIC is highest (0.51), and only 10 samples were misclassified (Table~\ref{tab:luke1} (b)). The model chosen by the LPBIC gave slightly worse classification performance ($\text{ARI}=0.47$ and 11 misclassifications; cf. Table~\ref{tab:luke1} (a) and (b)) .
The model selected by the BIC exhibits much worse classification performance, with an associated ARI of 0.29.

\section{Summary}\label{se:disc}
A model selection criterion, the ALPBIC, is proposed for mixture model selection in high-dimensional data applications. The ALPBIC is developed through a penalized likelihood-based approach, where the penalty term is akin to the adaptive LASSO. The ALPBIC is especially designed for high-dimensional data, where the BIC has known drawbacks. Through penalizing the likelihood by the adaptive LASSO, the resulting model selection criterion is consistent, which means it can consistently select the right number of components and covariance structure when $n \longrightarrow \infty$. The criterion also satisfies oracle property, which means the estimated parameters of the best model selected by the criterion satisfy consistency, sparsity, and asymptotic normality.

We used simulations to illustrate the consistency of the ALPBIC. These simulations also illustrate that while the LPBIC is a useful criteria for large $p$, it suffers from lack of consistency, i.e., it cannot consistently choose the same model for varying $n$. This is due to the fact that the LPBIC is derived through penalization of the likelihood where the penalty is the LASSO, which fails to satisfy oracle properties --- this drawback is resolved herein by using the adaptive LASSO. 

Although we used the PGMM family to illustrate our approach, the ALPBIC can be applied to any mixture model selection problem. 
A likelihood-based approach through penalization of the covariance parameters is a logical extension of the present work as it would lead to greater parsimony. The main obstacle in doing so is the complex sampling distribution of the off-diagonal elements of the component covariance matrices. We will aim to mitigate this problem by first extending the penalization to diagonal covariance matrices, before further extending this to a full matrix by imposing a Wishart distribution on its maximum likelihood estimator. 


\appendix
\section{Derivation of ALPBIC}
\label{se:app1}
APBIC is derived in a way similar to the derivation of LPBIC \citep{bhattacharya12}. To derive the ALPBIC, we have to maximize (\ref{eq:penlike}). Using (\ref{eq:penalty}), the second term becomes 
$$n\lambda_n\sum_{g=1}^G\int \pi_g \sum_{j=1}^{p}w_{gj}|\mu_{gj}|\mbox{d}\mathbf\pi_g=
n\frac{\lambda_n}{G}\sum_{g=1}^G \sum_{j=1}^{p_{g}}w_{gj}\left[|\hat\mu_{gj}|+\frac{1}{2}\frac{\mbox{sign}(\hat\mu_{gj})}{\hat\mu_{gj}}(\mu^2_{gj}-\hat\mu^2_{gj})\right],$$ where $p_g$ is the number of non-zero mean components in class~$g$.
Under the assumption made in Section~ \ref{se:method}, $\vecmu_g$ is at most $p_g$ dependent, and the Weak Law of Large Numbers holds. for $p$ large, $\sum_{g=1}^G p_g$ is a large number and so 
$$\frac{\sum_{g=1}^G \sum_{j=1}^{p_{g}}\left(\mu^2_{gj}-\hat\mu^2_{gj}\right)}{{\sum_{g=1}^G p_g}}  \overset{P}{\longrightarrow}\frac{\sum_{g=1}^G \sum_{j=1}^{p_{g}}\left(I(\hat{\mathbf\mu}_g)^{-1}\right)_{jj}}{{\sum_{g=1}^G p_g}}.$$
Thus the second term becomes $\frac{n\lambda}{G}\sum_{k=1}^G \sum_{j=1}^{p_g}w_{gj}\left[|\hat\mu_{gj}|+\frac{\left(I(\hat{\mathbf\mu}_g)^{-1}\right)_{jj}}{|\hat\mu_{gj}|}\right].$

Taylor's series expansion of the first term gives 
\begin{equation*}\begin{split}
\int \text{exp}&\left[\log \mathcal{L}(\varthet \mid \vecx)\mathcal{G}(\varthet)\right]\mbox{d}\mathbf\Theta\\&= \int \text{exp}\left[\log \mathcal{L}(\hat{\vecthet} \mid \vecx)\mathcal{G}(\hat{\varthet})+ (\varthet-\hat{\varthet})\frac{\partial \log \mathcal{L}(\varthet)\mathcal{G}(\varthet)}{\partial{\varthet}}-\frac{1}{2}(\varthet-\hat{\varthet})' \mathcal{H}_{\hat{\varthet}}(\varthet-\hat{\varthet})\right]\mbox{d}\varthet,
\end{split}\end{equation*}
where $\mathcal{H}$ is the second derivative matrix of $\log \mathcal{L}(\varthet)\mathcal{G}(\varthet)$. Since $\hat{\varthet}$ is derived by maximizing the penalized likelihood, the second term within the integral is $$(\varthet-\hat{\varthet})\frac{\partial\varphi_n(\vecmu)}{\partial{\varthet}},$$ where $\varphi_n(\vecmu)$ is the LASSO penalty function. Since $\hat\varthet$ is close to $\varthet$, using (\ref{eq:penalty}) and the mean-value theorem, the second term within the integral becomes 
$$\frac{n\lambda_n}{G}\sum_{g=1}^G \sum_{j=1}^{p_{g}}w_{gj}\mbox{sign}(\mu_{gj}).$$
The third term within the integral similarly is $$\frac{1}{2} (\tilde{\varthet}-\hat{\tilde{\varthet}})'\mathcal{H}_{\hat{\tilde{\varthet}}}(\tilde{\varthet}-\hat{\tilde{\varthet}}),$$
where $\tilde{\varthet}$ is the set of non-zero parameters and $\hat{\tilde{\varthet}}$ is their estimate. Using a Laplace approximation on $\mathcal{H}$ and applying the Weak Law of Large Numbers, as in the usual BIC, we arrive at $$\log \mathcal{L}(\hat{\varthet} \mid \vecx)-\frac{1}{2}\tilde \rho \log n,$$ where $\tilde \rho= \mbox{dim}(\hat{\tilde{\varthet}}) $. This, combined with the second term of (\ref{eq:penlike}), gives (\ref{eq:pbic}). 

\section{Proof of the Asymptotic Properties of the ALPBIC}\label{app:b} 

To prove Part a of the theorem in Section \ref{asymp}, let $r_n= n^{-{1/2}}\sqrt{\log n}$. It is sufficient to prove that for any given $\epsilon >0$, there exists a constant $M_{\epsilon}$ such that $$\underset{n \rightarrow \infty}{\text{lim}}\text{P}\left( \underset{\mid\mid \matu \mid\mid= M_{\epsilon}}{\text{sup}}\text{ALPBIC}\left(\varthet_0+r_n\matu\right)< \text{ALPBIC}\left(\varthet_0\right)\right) \geq 1-\epsilon .$$ Thus, we have to show that, with large probability, there is a local maximum inside $$\left\lbrace \varthet_0+r_n\matu : \mid\mid \matu \mid\mid  \leq M_{\epsilon} \right\rbrace.$$
Using (\ref{eq:pbic}), we can write 
\begin{equation}\begin{split}\label{eq:asymp}
&\text{ALPBIC}\left(\varthet_0+r_n\matu\right)- \text{ALPBIC}\left(\varthet_0\right) \approx
2l(\varthet_0+r_n\matu \mid \vecx)-2l(\varthet_0 \mid \vecx)
-\left(\tilde \rho_1-\tilde \rho_2\right)\log n \\
&\qquad-2n \frac{\lambda_n}{G} \sum_{g=1}^G\sum_{j=1}^{p_g} w_{gj}\left[\mid \mu_{gj}+r_n u_{0gj}\mid -\mid \mu_{0gj} \mid + \frac{\left(I(\vecmu_{0g}+r_n\matu)^{-1}\right)_{jj}}{|\mu_{0gj}+r_n u_{gj}|}-\frac{\left(I(\vecmu_{0g})^{-1}\right)_{jj}}{|\mu_{0gj}|}\right],
\end{split}\end{equation}
where $l(\varthet_0 \mid \vecx)=\log \mathcal{L}(\varthet_0 \mid \vecx)$, $\tilde \rho_1$ and $\tilde \rho_2$ are the number of non-zero elements in $\varthet_0+r_n\matu$ and $\varthet_0$, respectively, $u_{gj}$ is the $j$th element of $\matu$ corresponding to the $g$th component, and $p_g$ is the number of non-zero elements in $\vecmu_{0g}$.
If $p^1_g$ is the number of non-zero elements in $\vecmu_{0g}+r_n\matu$, then $$2n \frac{\lambda_n}{G} \sum_{g=1}^G\sum_{j=p^1_g}^{p_g} w_{gj}
\left[\mid \mu_{0gj}+r_n u_{gj}\mid -\mid \mu_{0gj} \mid + \frac{\left(I(\vecmu_{0g}+r_n\matu)^{-1}\right)_{jj}}{|\mu_{0gj}+r_n u_{gj}|}-\frac{\left(I(\vecmu_{0g})^{-1}\right)_{jj}}{|\mu_{0gj}|}\right]$$ is negligible in order comparison, as we shall see later in the proof. Also, for sufficiently large $n$, we can make $\text{sign}\left(\mu_{0gj}+r_nu_{gj}\right)-\text{sign}\left(\mu_{0gj}\right)=0$.

Using Taylor's expansion, we can write  $$l(\varthet_0+r_n\matu \mid \vecx)-l(\varthet_0 \mid \vecx)= n^{-{1/2}}(\log n)^{1/2} l'(\varthet_0)' \matu-\frac{\log n}{2}\left(\matu' I(\varthet_0)\matu\right)(1+o(1)).$$
Also, for sufficiently large $n$, $\tilde\rho_1-\tilde\rho_2 = o(1)$. For the third term of (\ref{eq:asymp}), using the assumption $(\log n)^{1/2}n^{\kappa+1/2}\lambda_n\longrightarrow 0$, we have
\begin{equation}\begin{split}
 n\frac{\lambda_n}{G}& \sum_{g=1}^G\sum_{j=1}^{p_g} w_{gj}\left[|\mu_{0gj}+r_n u_{gj}| -|\mu_{0gj}|\right] \longrightarrow  (n\log n)^{1/2} \frac{\lambda_n}{G} \sum_{g=1}^G\sum_{j=1}^{p_g} |\mu_{0gj}|^{-\gamma}u_{gj}\text{sign}(\mu_{0gj})\nonumber
  \end{split}\end{equation}
because $\tilde\mu_{gj}$ is a $\sqrt{n}$-consistent estimator and hence $w_{gj} \rightarrow |\mu_{0gj}|^{-\gamma} $. On the other hand, $\lambda_n w_{gj}= n^{\gamma/2}\lambda_n |n^{{1/2}} \tilde\mu_{gj}|^{-\gamma}$, where $\tilde\mu_{gj}$ is a $\sqrt{n}$-consistent estimator, and hence $n^{{1/2}} \tilde\mu_{gj} = \mathcal{O}(1)$. Using this result, 
\begin{equation}
n \frac{\lambda_n} {G} \sum_{g=1}^G\sum_{j=1}^{p_g} w_{gj}\left| \frac{\left(I(\vecmu_{0g}+r_n\matu)^{-1}\right)_{jj}}{|\mu_{0gj}+r_n u|}-\frac{\left(I(\vecmu_{0g})^{-1}\right)_{jj}}{|\mu_{0gj}|} \right|\\ \nonumber
\leq n^{(\gamma+2\kappa+1)/2}(\log n)^{1/2}\lambda_n M,
\end{equation}\nonumber
where $$M= n^{1/2}(\log n)^{-{1/2}}\underset{g,j}{\max}|\sqrt{n} \tilde\mu_{gj}|^{-\gamma}\frac{\underset{\varthet : ||\varthet-\varthet_0||= \mathcal{O}\left( r_n\right)}{\text{max}} \left(I(\mu_{gj})^{-1}\right)_{jj}}{\underset{\varthet : ||\varthet-\varthet_0|| = \mathcal{O}\left( r_n\right)}{\min}|\mu_{gj}|}.$$ From the regularity conditions, $l'(\varthet_0)=\mathcal{O}\left(n^{1/2}\right) $ and $$\underset{g,j}{\max}\left(I(\mu_{0gj}+r_n\matu)^{-1}\right)_{jj}= \mathcal{O}\left(n^{-{1/2}}\right).$$ Therefore, $M \rightarrow 0$ as $n \rightarrow \infty$. 

Because $$\frac{n^{(\gamma+2\kappa+1)/2}\lambda_n}{(\log n)^{1/2}}
=\mathcal{O}(1) \quad\text{ and }\quad \frac{n^{(2\kappa+1)/2}\lambda_n}{(\log n)^{1/2}}=o(1),$$ by order comparison, we conclude that $-{\log n}/2\left(\matu' I(\varthet_0)\matu\right)
(1+o(1))$ is the sole leading term of (\ref{eq:asymp}). Because the information matrix is positive definite, $$\underset{n \rightarrow \infty}{\text{lim}}\text{P}\left( \underset{\mid\mid \matu \mid\mid= M_{\epsilon}}{\text{sup}}\text{ALPBIC}\left(\varthet_0+r_n\matu\right)< \text{ALPBIC}\left(\varthet_0\right)\right) \geq 1-\epsilon,$$ and this concludes the proof of Part~a.

To prove Part~b, use the mean-value theorem,
$$
l(\hat\varthet_{1}, \hat\varthet_{2} \mid \vecx)-l(\hat\varthet_{1}, \mathbf{0}\mid \vecx)
=\left[\frac{\partial l(\hat\varthet_{1}, \mathbf{\xi})}{{\partial \varthet_{2}}}\right]'\hat\varthet_{2},$$
where $\mid\mid \mathbf{\xi} \mid\mid \leq \mid\mid \hat\varthet_{2} \mid\mid = \mathcal{O}\left(n^{-{1/2}}(\log n)^{1/2}\right).$ Also,  
\begin{equation}\begin{split}
\left| \left| \frac{\partial l(\hat\varthet_{1}, \mathbf{\xi})}{\partial \varthet_{2}}-
\frac{\partial l\left(\varthet_{0}, \mathbf{0}\right)}{\partial \varthet_{2}} \right|\right| \leq & \left|\left| \frac{\partial l(\hat\varthet_{1}, \mathbf{\xi})}{\partial \varthet_{2}}-\frac{\partial l(\hat\varthet_{1}, \mathbf{0})}{\partial \varthet_{2}} \right|\right|+\left|\left| \frac{\partial l(\hat\varthet_{1}, \mathbf{0})}{\partial \varthet_{2}}-\frac{\partial l\left(\varthet_{0}, \mathbf{0}\right)}{\partial \varthet_{2}} \right|\right|\\ \nonumber
\leq & \sum_{i=1}^n M_2(z_i) \left[\left|\left|\mathbf{\xi}\right|\right|+ \left|\left| \hat\varthet_{1}-\varthet_0\right|\right|\right] =\mathcal{O}\left((n\log n)^{1/2}\right). \nonumber
\end{split}\end{equation} 
From the regularity conditions, $\partial l\left(\varthet_{0}, \mathbf{0}\right)/{\partial \varthet_{2}}$ is of order $\mathcal{O}\left(n^{1/2}\right)$ and so $$\frac{\partial l(\hat\varthet_{1}, \mathbf{\xi})}{{\partial \varthet_{2}}}=\mathcal{O}\left((n\log n)^{1/2}\right).$$ Therefore, from these order assessments, we conclude that $$l\left(\hat\varthet_{1}, \hat\varthet_{2}\right)-l\left(\hat\varthet_{1}, \mathbf{0}\right) = \mathcal{O}\left((n\log n)^{1/2}\right)\sum_{g=1}^G\sum_{j=p_g+1}^p \hat\mu_{gj},$$ where $p_g$ is defined as in (\ref{eq:pbic}). 
Also, $(\tilde\rho_1-\tilde\rho_2)\log n= \sum_{g=1}^G\left(p-p_g\right)\log n$. Hence,
\begin{equation}\begin{split}\label{eq:sparsity1}
&\text{ALPBIC}\left(\hat\varthet_{1}, \hat\varthet_{2} \right)- \text{ALPBIC}\left(\hat\varthet_{1}, \mathbf{0}\right) \approx \sum_{g=1}^G\sum_{j=p_g+1}^p \left[ \mathcal{O}\left((n\log n)^{1/2}\right)\hat\mu_{gj} \right.\\ 
&\left.\qquad\qquad\qquad - 2n\frac{\lambda_n}{G}  w_{gj}\left(\mid \hat \mu_{gj} \mid + \frac{\left(I(\hat{\vecmu}_g)^{-1}\right)_{jj}}{|\hat\mu_{gj}|}-\mbox{sign}\left(\hat\mu_{gj}\right) \right)\right] -\sum_{g=1}^G\left(p-p_g\right)\log n. 
 \end{split}\end{equation}
Because $\mid\mid \hat\varthet_2\mid\mid = \mathcal{O}\left(n^{-{1/2}}(\log n)^{1/2}\right),$  we can conclude that 
\begin{equation}\begin{split}
&\left|\sum_{g=1}^G\sum_{j=p_g+1}^p  \mathcal{O}\left((n\log n)^{1/2}\right)\hat\mu_{gj} \right| = \mathcal{O}\left(\log n\right),\\
&\sum_{g=1}^G\sum_{j=p_g+1}^p n\frac{\lambda_n}{G}  w_{gj}\mid \hat \mu_{gj} \mid /{\log n} \longrightarrow 0, \qquad\text{and} \\
&\sum_{g=1}^G\sum_{j=p_g+1}^p n\frac{\lambda_n}{G}  w_{gj} \left(I(\hat{\vecmu}_g)^{-1}\right)_{jj}/{|\hat\mu_{gj}|}  \longrightarrow 0
\end{split}\end{equation}
 with probability tending to 1. This is due to assumption (\ref{assump:3}) and the fact that $$n^{(\gamma+1)/2}\lambda_n/{(\log n)^{1/2}} \longrightarrow 0.$$ Also, $$\left|\sum_{g=1}^G\sum_{j=p_g+1}^p n\frac{\lambda_n}{G}  w_{gj}  \text{sign}\left(\hat\mu_{gj}\right)\right|\frac{1}{\log n} \leq \mathcal{O}(1)\left(n/{\log n}\right)^{1/2}n^{{\gamma+1}/2}\lambda_n \sum_{g=1}^G(p-p_g).$$
Therefore, by order assessment, as $n \rightarrow \infty$, \eqref{eq:sparsity1} boils down to the leading term 
\begin{equation}\begin{split}
\sum_{g=1}^G\sum_{j=p_g+1}^p \left[\mathcal{O}\left((n\log n)^{1/2}\right)\hat\mu_{gj}-- 2n\frac{\lambda_n}{G}  w_{gj}\text{sign}(\hat\mu_{gj}) \right]-\sum_{g=1}^G\left(p-p_g\right)\log n \\
= \sum_{g=1}^G\sum_{j=p_g+1}^p \left[ \mathcal{O}\left((n /\log n)^{1/2}\right)\left(\hat\mu_{gj}-o(1)\right)-1\right],
\end{split}\end{equation}
 which is negative for sufficiently large $n$ with probability tending to 1. This concludes the proof of Part~b.


The proof of Parts c and d follow in the fashion described by \cite{khalili}. 
\end{document}